# Models of Manipulation on Aggregation of Binary Evaluations

Elad Dokow*    Dvir Falik†

September 21, 2018

**Abstract**

We study a general aggregation problem in which a society has to determine its position on each of several issues, based on the positions of the members of the society on those issues. There is a prescribed set of feasible evaluations, i.e., permissible combinations of positions on the issues. Among other things, this framework admits the modeling of preference aggregation, judgment aggregation, classification, clustering and facility location. An important notion in aggregation of evaluations is strategy-proofness. In the general framework we discuss here, several definitions of strategy-proofness may be considered. We present here 3 natural *general* definitions of strategy-proofness and analyze the possibility of designing an annonymous, strategy-proof aggregation rule under these definitions.

## 1 Introduction

There is, by now, a significant body of literature on the problem of aggregating binary evaluations. A society has to determine its position (yes/no) on each of several issues, based on the positions of the members of the society on those issues. There is prescribed set $X$ of feasible evaluations, i.e., permissible combinations of positions on the issues ($X$ may be viewed as a subset of $\{0,1\}^m$, where $m$ is the number of issues). The members of the society report their opinion to an aggregation mechanism, called the *aggregator*, which outputs society's aggregated opinion. Many examples include preference aggregation (where the issues are pairwise comparisons and feasibility reflects rationality), and judgment aggregation (where the issues are logical propositions and feasibility reflects consistency) can be presented by this framework, we will present some of them in subsection 1.1. We shall refer to this framework as Judgment Aggregation throughout the paper, as this model is actually as general as the entire framework.

This paper deals with introducing a general definition of manipulations and strategy-proofness to this model. Generally speaking, we assume that each member of the society has some preference over the possible outcomes, which is derived from her true opinion on the issues. Under this assumption, it may not always be the rational course of action for a member of the society to report her true opinion to the aggregator. Such an occurence is called a *manipulation* of the aggregator. An aggregator which is immune to manipulations is called *strategy-proof*. There is no canonical way to define the concept of manipulation in judgment aggregation. In this paper we wish to initiate a systematic study of the range of general definitions of manipulations for Judgment aggregation and analyze the possibility of designing strategy-proof aggregators for given evaluation spaces under a given definition of manipulation.

*Economics Department, Bar-Ilan University. eladokow@gmail.com
†School of Mathematical Sciences, Tel Aviv University.

The field of judgment aggregation was inspired by the classical impossibility theorem of Arrow [Arr63] on preference aggregation. Arrow's theorem shows that it is impossible to design an aggregation aggregator for preference aggregation that satisfies several natural properties. Preference aggregation is an instance of judgment aggregation, and much of the work done in this context related to the generalization of Arrow's theorem to Judgment aggregation.

Another classical result that is known to be connected to Arrow's theorem is Gibbard-Satterthwaite's theorem [Gib73, Sat75]. This theorem deals with a variant of preference aggregation called social choice, where preferences are aggregated into a single choice, and shows the impossibility of designing strategy-proof aggregators in that context. This is the work that inspired the line of research we present here. Since social choice is *not* an instance of judgment aggregation, there is a degree of freedom in the choice of definition for the analogue of manipulation for judgment aggregation. The results of Arrow and Gibbard-Satterthwaite and their connection to judgment aggregation will be introduced and explained in subsection 1.2.

## 1.1 Judgment Aggregation Examples

In this subsection we present a few applications taken from several distinct areas, that can all be modeled via the judgment aggregation framework:

- **Preference Aggregation:** In this setting, the society wishes to rank $k$ alternatives, in order of preference, where each voter has its own private order of preference. This problem has been studied since the days of the French revolution, by the Marquis de Condorcet. We will only address in this paper the case where the ranking has to be full - i.e. there is always a strict preference between 2 alternatives[1]. The set of issues is the set of pairwise preferences between every 2 alternatives, so $m = \binom{k}{2}$. Each pair of alternatives may have 2 possible preferences, so the full opinion has a binary encoding. The permissible evaluations are the preferences that encode a full transitive order.

    For example, when $k = 3$, the set of alternatives is $\{a, b, c\}$, the set of issues is $\{a \succ b, b \succ c, c \succ a\}$, and the permissible evaluations are all possible evaluations except 000 and 111, which encode a non-transitive order.

    Condorcet noticed that a specific natural aggregator, that chooses in each issue the majority opinion of the society in that issue, does not always produce permissible evaluations. This is known as "Condorcet's Paradox".

    Condorcet's Paradox motivated the study of social choice theory, beginning in Arrow's theorem [Arr63].

---

[1]There are works that deal with the more general framework, where the preferences are not strict, see, e.g. [Arr63].

|         | $a \succ b$ | $b \succ c$ | $c \succ a$ |
|---------|-------------|-------------|-------------|
| Voter 1 | 1           | 1           | 0           |
| Voter 2 | 0           | 1           | 1           |
| voter 3 | 1           | 0           | 1           |
| Aggr.   | 1           | 1           | 1           |

Table 1: Condorcet's paradox

- **Judgment Aggregation:** In the last decade, there is a growing body of work in the field of judgment Aggregation, where judges need to come to a decision on a set $J$ of connected issues. The connection between the issues is expressed by a set of permissible evaluations $X \subseteq \{0,1\}^J$. The canonical example in this context is the *doctrinal paradox* (also called *the discursive dilemma* ), in which a court has to decide whether a defendant is guilty. In order to declare him guilty, they must hold the opinion that he has committed the crime and that he was sane at the time. The set of permissible evaluation, therefore, is

$$X = \{(p,q,r) | r = p \wedge q\}$$

The so called "paradox" arises when a majority of the judges think that the defendant has committed the crime, and a majority of the judges believe he was sane at the time, but only a minority of the judges believe both to hold.

|          | Murdered | Sane | Guilty |
|----------|----------|------|--------|
| Judge 1  | 0        | 1    | 0      |
| Judge 2  | 1        | 0    | 0      |
| Judge 3  | 1        | 1    | 1      |
| Majority | 1        | 1    | 0      |

Table 2: Doctrinal Paradox

Many works done in recent years discussed this general framework[2]. In particular, the conditions on $X$ for which Arrow's theorem holds has been extensibly studied[3].

- **Classification**[4]**:** A set of $m$ points has to be classified, and there is a prescribed set of classifiers. For instance, consider the case where the points lie in $\mathbb{R}^k$, and the classifiers are all the linear half-spaces. The society is composed of $n$ agents, each has its own classification of the points, and the aggregator must select a classifier based on the opinions of the agents.

  This problem fits into our framework when the classifiers are encoded as the vector of their classification of all the points.

  For example, consider the points to be $\{(0,0),(0,1),(1,0),(1,1)\}$. The possible linear classifiers in this case are all classifiers except for 0110 and 1001:

$$X = \{0,1\}^4 \setminus \{0110, 1001\}$$

---

[2]See the survey [LP10].
[3]See [NP10, DH10]
[4]See [RMR09]

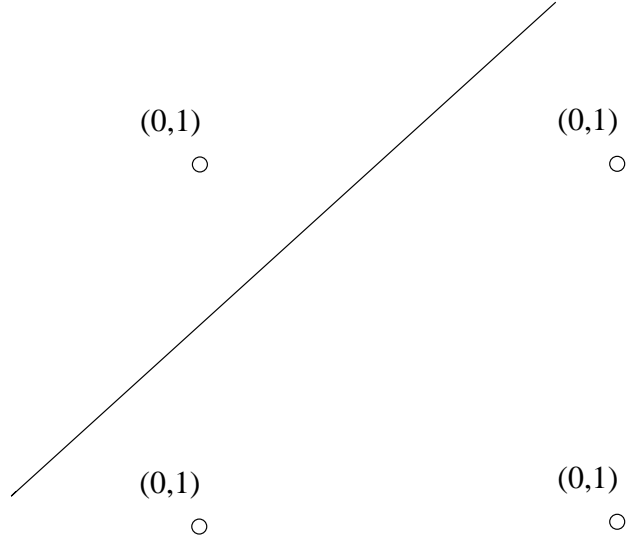

Figure 1: Classification example

- **Facility Location:** In this problem, an aggregator is given $k$ points in some metric space, and is required to choose a location for a facility that services all these points. The location of each point is reported to the aggregator by a single agent, which may or may not be truthful. The aggregator should optimize the distance of the chosen location from the locations of the points.

  We can encode this problem under our framework in the case that the metric space that is used is isomorphic to an induced subgraph of the Boolean hypercube equipped with the Hamming metric. The set $X$ of permissible evaluations will be the set of vertices in the Boolean hypercube corresponding to the given metric space.

  For instance, a simple cycle on $2m$ vertices can be encoded as

  $$X = \left\{1^i 0^{m-i} | i \in \{0..m\}\right\} \cup \left\{0^{m-i} 1^i | i \in \{1..m-1\}\right\}$$

## 1.2 Introduction to Social Choice

Condorcet's paradox motivated the study of Social Choice Theory, beginning in Arrow's impossibility theorem. Relating to preference aggregation, Arrow's theorem states that it is impossible to design a social aggregator that satisfies some natural conditions. It is natural to assume that for every social aggregator, the aggregated order is always a transitive order (consistent), that it agrees with a unanimous vote (Pareto optimal), and that it is influenced by the opinions of more than one voter (non-dictatorial). Condorcet's proposed aggregator satisfied the property that its decision on the social preferences between alternatives $a$ and $b$ depends only on the individual preferences between $a$ and $b$. This property is known as Independence of Irrelevant alternatives, or IIA. Arrow showed that a social aggregator on 3 or more alternatives that satisfies IIA, cannot be consistent, Pareto-optimal and non-dictatorial.

When addressing the general framework of judgment aggregation, IIA has a natural generalization - that the decision of the aggregator in each issue depends only on the individual opinions on that issue. Wilson [Wil75] was the first to propose this general framework in connection with extending Arrow's result to it. Later on, this scheme was restudied in the context of judgment aggregation [LP02], where the doctrinal paradox was first introduced. Much of the work in this field concentrated on characterizing evaluation spaces $X$ for which Arrow's theorem holds.

A variant of preference aggregation is *social choice*, where the social aggregator is only required to choose society's preferred alternative, instead of producing a full ranking of the alternatives. Gibbard and Satterthwaite [Gib73, Sat75] showed an impossibility theorem for social choice aggregators. Their theorem deals with the game-theoretic notion of *strategy-proofness*. An aggregator is called *strategy-proof* if it does not give the voters an incentive to lie about their preference when reporting it to the aggregator. A situation when a voter gains (makes society's opinion better according to her true opinion) by lying is called a manipulation. The theorem states that there is no non-dictatorial social aggregator that is strategy-proof (if the number of alternatives that can win the election is at least 3).

The theorems of Arrow and Gibbard-Satterthwaite are known to be highly connected. We will elaborate further on the strong connection between the properties of IIA in preference aggregation and strategy-proofness in social choice.

In contrast with IIA, the concept of manipulation introduced in Gibbard-Satterthwaite's theorem does not have a canonical extension in the judgment aggregation model.

In the social choice setting, the opinion of a voter determines her preferences over all the alternatives. If a voter misreports her opinion, thus changing the identity of the chosen alternative from $a$ to $b$ (assuming the other votes do not change), it is clear which of the two results is preferred by her. If $b$ is preferred by her over $a$, this is a manipulation of the election mechanism.

In the general setting of judgment aggregation, the preference of a voter over all possible results is not clear from her opinion, and each problem can have a different interpretation of this notion.

Another difference between social choice and the general setting is that in social choice the space of opinions of the society and the space of possible results of the election process are different.

In order to reach a general definition of manipulation, we assume that each voter desires the aggregated evaluation to agree with her personal evaluation in all or some of the issues. Since there may be situations where some of the issues change for the better and some for the worse (in the manipulator's view), there is still a degree of freedom in the choice of a definition of a manipulation.

This work discusses 3 natural definitions of the concept of manipulation on judgment aggregation. One of the definitions was defined and discussed in [NP10, DL07], and it leads to impossibility results similar to those that were mentioned here. The other two definitions allow non-dictatorial aggregators, and we will discuss the construction of such aggregators in the general case.

## 1.3 Example of manipulations

Consider an aggregator for preference aggregation on 3 alternatives, that uses the *plurality* method. It selects the ranking that was voted for the highest number of times. In case of a tie, it uses a lexicographical order to choose the ranking. Consider the following profile:

|         | $a \succ b$ | $b \succ c$ | $c \succ a$ |
|---------|-------------|-------------|-------------|
| Voter 1 | 1           | 1           | 0           |
| Voter 2 | 0           | 1           | 1           |
| voter 3 | 1           | 0           | 1           |
| Aggr.   | 1           | 1           | 0           |

Table 3: Preference aggregation by plurality - Profile 1

The second voter has the society agreeing with her on the second issue - $b \succ c$, and disagreeing with her in the other issues. She can change this when reporting a different opinion -

|         | $a \succ b$ | $b \succ c$ | $c \succ a$ |
|---------|-------------|-------------|-------------|
| Voter 1 | 1           | 1           | 0           |
| Voter 2 | 1           | 0           | 1           |
| voter 3 | 1           | 0           | 1           |
| Aggr.   | 1           | 0           | 1           |

Table 4: Preference aggregation by plurality - Profile 2

Now the society agrees with her original opinion in the third issue - $c \succ a$, and disagrees with her on the other issues.

If her main interest was in getting society to agree with her in the third issue, then she has successfully manipulated the aggregator.

If her main interest was to get the society to agree with her in as many issues as possible, then she has not manipulated the aggregator, as in both cases the aggregated opinion agreed with her in only 1 issue.

Here is a different scenario -

|         | $a \succ b$ | $b \succ c$ | $c \succ a$ |
|---------|-------------|-------------|-------------|
| Voter 1 | 1           | 0           | 1           |
| Voter 2 | 0           | 1           | 1           |
| voter 3 | 0           | 1           | 0           |
| Aggr.   | 1           | 0           | 1           |

Table 5: Preference aggregation by plurality - Profile 3

The second voter has the society agreeing with her on the third issue - $c \succ a$, and disagreeing with her in the other issues. When she reports a different opinion -

|         | $a \succ b$ | $b \succ c$ | $c \succ a$ |
|---------|-------------|-------------|-------------|
| Voter 1 | 1 | 0 | 1 |
| Voter 2 | 0 | 1 | 0 |
| voter 3 | 0 | 1 | 0 |
| Aggr.   | 0 | 1 | 0 |

The society now agrees with her original opinion on the first 2 issues. She has gained in the number of issues the society agrees with her. However, she has lost the agreement with the society on the third issue.

The third and final scenario is:

|         | $a \succ b$ | $b \succ c$ | $c \succ a$ |
|---------|-------------|-------------|-------------|
| Voter 1 | 1 | 0 | 1 |
| Voter 2 | 0 | 1 | 1 |
| voter 3 | 0 | 0 | 1 |
| Aggr.   | 1 | 0 | 1 |

The second voter has the society agreeing with her on the third issue - $c \succ a$, and disagreeing with her in the other issues. When she reports a different opinion -

|         | $a \succ b$ | $b \succ c$ | $c \succ a$ |
|---------|-------------|-------------|-------------|
| Voter 1 | 1 | 0 | 1 |
| Voter 2 | 0 | 0 | 1 |
| voter 3 | 0 | 0 | 1 |
| Aggr.   | 0 | 0 | 1 |

The society now agrees with her original opinion on the first and last issues. She has gained agreement in the first issue and did not lose agreement on any of the other issues.

When designing an aggregator, we need to know what type of manipulations we wish to be immune against. A maximal requirement is to be immune from manipulations that gain in any of the issues (we will call these *partial manipulations*). A minimal requirement is immunity from manipulations that don't lose agreement in any of the issues (we will call these *full manipulations*). There could be other types of manipulations in between, for instance, manipulations that gain in the number of issues agreed with the society (we will call these *Hamming manipulations*). For the particular example discussed here, we will show that we are able to design a non-dictatorial aggregator that avoids manipulations of the second and third type.

## 1.4 Structure of the paper and results

### 1.4.1 Structure

In Chapter 2. we present the formal model used throughout the paper. We then dedicate a chapter for each of the 3 types of manipulations mentioned above. Chapter 3 dicusses the *partial* *manipulation*, chapter 4 discusses the *full manipulation*, and chapter 5 discusses the *Hamming manipulation*.

### 1.4.2 Results

- **Partial manipulation:** Partial manipulation was already discussed in previous works. We bring the results here for completion. These results characterize evaluation spaces $X$ for which the only partial manipulation free aggregators are dictatorial. These results are based on the connection between partial manipulations and IIA.

- **Full manipulation** We show that there is a family of non-dictatorial full manipulation free aggregators for every evaluation space $X$. In addition, for every evaluation space $X$ some members of this family are also Hamming manipulation free.

  We next turn to the question of annonymous full manipulation free aggregators. For every evaluation space $X$, we construct a family of aggregators that are annonymous and full manipulation free. These aggregators are also "close" to being partial manipulation free in some sense.

  We also show that when the welfare of a voter is defined as the Hamming distance between its opinion and society's desicion, the social welfare maximizer is a full manipulation free aggregator.

- **Hamming manipulation** Again, we discuss the possibility of constructing a *annonymous* Hamming manipulation free aggregator. Since every *Hamming manipulation free* aggregator is also a *full manipulation free* aggregator, we try and characterize the evaluation spaces $X$ for which the full manipulation free annonymous aggregators mentioned above are also Hamming manipulation free.

  We do not have a full characterization of these aggregators. We describe some conditions that affect the Hamming strategy proofness of these aggregators, based on the geometry of the evaluation space.

  We apply these techniques to demostrate that in the case of preference aggregation on 3 alternatives these aggregators are Hamming manipulation free, and for 4 alternatives we show that a subfamily of these aggregators are not Hamming manipulation free.

## 2 The setting

We consider a finite, non-empty set of issues $J$. For convenience, if there are $m$ issues in $J$, we identify $J$ with the set $\{1,...m\}$ of coordinates of vectors of length $m$. A vector $x = (x_1,...,x_m) \in \{0,1\}^m$ is an *evaluation*. We assume that some non-empty subset $X$ of $\{0,1\}^m$ is given. The evaluations in X are called feasible, the others are infeasible. We shall also use this terminology for partial evaluations: for a subset of issues $K$, a $K$-evaluation is feasible if it lies in the projection of $X$ on the coordinates in $K$, and is infeasible otherwise. A *society* is a finite, non-empty set $N$. For convenience, if there are $n$ individuals in $N$, we identify $N$ with the set $\{1,...,n\}$. If we specify a feasible evaluation $x^i = (x_1^i,...,x_m^i) \in X$ for each individual $i \in N$, we obtain a profile of feasible evaluations $\mathbf{x} = (x_j^i) \in X^n$. We may view a profile as an $n \times m$ matrix all of whose rows lie in X. We use superscripts to indicate individuals (rows) and subscripts to indicate issues (columns). An *aggregator* for N over X is a mapping $f : X^n \to X$. It assigns to every possible profile of individual feasible evaluations, a social evaluation which is also feasible. Any aggregator f may be written in the form $f = (f_1, ..., f_m)$ where $f_j$ is the j-th component of f. That is, $f_j : X^n \to \{0, 1\}$ assigns to every profile the social position on the j-th issue. By abuse of notation, for a profile $\mathbf{x} \in X^n$, we will write $\mathbf{x} = (x^i, x^{-i})$, where $x^{-i} \in X^{n-1}$ are the evauations of all individuals except for the $i$'th individual.

**Definition 2.1: Independency of Irrelevant Alternatives** An aggregator is called *independent of irrelevant alternatives* (IIA) if the society's position on any given issue depends only on the individual positions on that same issue.

$$\forall x, y \in X^n, j \in J, (x_j = y_j) \Rightarrow (f_j(\mathbf{x}) = f_j(\mathbf{y}))$$

**Definition 2.2: Annonimity** An aggregator is called *annonymous* if it does not depend on the names of the evaluators, i.e. for every permutation $p$ of the evaluators, $f(x^{p(1)}, ...x^{p(n)}) = f(x^1, ...x^n)$. ∎

**Definition 2.3: Monotonicity** An aggregator is called *monotone* if changing an individual's position on some issue never results in a change of the society's position on that issue in the opposite direction. ∎

**Definition 2.4: Dictatorship** An aggregator is called *dictatorial* if it obeys the opinion of only one of the evaluators - There exists an evaluator $i \in N$ such that $f(\mathbf{x}) = x^i$. ∎

## 2.1 Strategic voting and strategy-proofness

We now assume that each voter desires the aggregtaed evaluation to agree with her personal evaluation in all or some of the issues. Under this assumption, it may not always be the rational choice for the voter to declare her true evaluation to the aggregator. Given that the other voters voted in a specific way, lying about her evaluation may change society's position on certain issues to match hers, making society's position "closer" to hers, under some definition of closeness. (This is called *strategic voting*.) An evaluator $i$ is said to have a manipulation of an aggregator $f$ in the profile $\mathbf{x} \in X^n$ if she can report a false evaluation $y$ in a way that the resulting aggregated evaluation $w = f(y, x^{-i})$ is preferred by her, according to her opinion $x^i$, over the original aggregated evaluation $z = f(x^i, x^{-i})$. $y$ is called a manipulation of $i$ over $\mathbf{x}$.

What is left to decide is, when $w$ is preferred over $z$, according to $x^i$. For an issue $j \in J$, if $w_j = x_j^i \neq z_j$, we shall call $w$ *j-preferrable* over $z$ according to $x^i$. If $w_j = z_j$, then we say that $w$ and $z$ are *j-indifferent* to each other according to $x^i$. It is natural to assume that under any definition of manipulation, if $y$ is a manipulation of $i$ over $\mathbf{x}$, then there must be at least one issue $j \in J$ such that $w$ is $j$ preferrable over $z$ according to $x^i$. It is also natural to assume that under any definition of manipulation, if for every $j \in J$, $z$ is $j$-indifferent to $w$ or $j$-preferrable over $w$ according to $x^i$, then $y$ is not a manipulation of $i$ over $\mathbf{x}$.

We will base our definitions of manipulation on these assumptions.

**Definition 2.5: Partial Manipulation:** If there exists an issue $j \in J$ such that $w$ is $j$ preferrable over $z$ according to $x^i$, then $y$ is a *partial manipulation* of $i$ over $\mathbf{x}$. ∎

**Definition 2.6: Full Manipulation:** If there exists an issue $j \in J$ such that $w$ is $j$ preferrable over $z$ according to $x^i$, and for every issue $j' \in J$, $w$ is $j'$-preferred over $z$ or $j'$-indifferent to $z$ according to $x^i$, then $y$ is a *full manipulation* of $i$ over $\mathbf{x}$. ∎

All possible definitions of manipulations that fit our assumptions lie between these two definitions. A natural and interesting choice of a definition of manipulation is based on the (Weighted) Hamming metric. For two vectors $x, y \in \{0,1\}^m$, we define their weighted Hamming distance with weight $\omega \in \mathbb{R}_+^m$ where $\sum_{j=1}^m \omega_j = 1$ as

$$d_w(x,y) = \sum_{j=1}^m \omega_j |x_j - w_j|$$

.

We will deal with the case when all the voters share the same weight function $\omega$ of the issues, and use the following definition:

**Definition 2.7: Weighted Hamming Manipulation:** If $d_\omega(x^i, w) < d_\omega^j(x^i, z)$, then $y$ is a $\omega^j$-*Hamming manipulation* of $i$ over $\mathbf{x}$ . ∎

If $\omega$ is uniform over the issues, we omit it from the notation:

**Definition 2.8: Hamming Manipulation:** If $d(x^i, w) < d(x^i, z)$, where $d(x^i, w) = \sum_{i=1}^m |x_i - w_i|$ then $y$ is a *Hamming manipulation* of $i$ over $\mathbf{x}$ . ∎

In subsection 1.3, the first example was a partial manipulation which was not a Hamming nor full manipulation, and the second example was a partial manipulation and a Hamming manipulation, but not a full manipulation. The third example is of a full manipulation. A manipulation of any type is also a partial manipulation, and a full manipulation is also a manipulation of any other type.

The general definition of Partial manipulation has been studied in [NP10, DL07]. The Hamming manipulation has been studied for a specific instance of classification in [RMR09]. A geometric definition similar to the Hamming manipulation has been studied in the context of facility location ([NAT10]).

# 3 Partial Manipulation

## 3.1 Motivation

When there are no assumptions on the preferences of voters, we fear that any possible type of manipulation may be considered profitable by any of the voters. In that case, a strategy-proof aggregator must be partial-manipulation-free (PMF), as every manipulation is a partial manipulation. This definition of manipulation and the results in this section were introduced and discussed in previos works of Nehring and Puppe [NP10] and Deitrich and List [DL07].

Since this is the broadest definition of manipulation, being immune to it is difficult, and the main results are impossibility theorems regarding the construction of PMF aggregators.

## 3.2 Impossibilty Theorem

The property of being PMF gives rise to impossibility theorems under certain conditions on $X$, due to its connection to the property of being IIA

**Theorem 3.1:** *[NP10]*

*For all nonempty evaluation spaces $X \subseteq \{0,1\}^m$, an aggregator $f : X^n \to X$ is PMF if and only if it is IIA and monotone[5].*

**Proof:** It is trivial to see the non-monoticity implies partial-manipulability, and that IIA and monotinicity together imply non-manipulability. We will show that partial-manipulation-free aggregators are IIA.

Assume $f$ is not IIA - there exist two inputs $\mathbf{x}, \mathbf{y} \in X^n$ and an issue $j \in J$ such that $x_j = y_j$ and $f_j(\mathbf{x}) \neq f_j(\mathbf{y})$. W.l.o.g., we may assume that there exists exactly one voter $i \in N$ such that $\mathbf{x}^i$ and $\mathbf{y}^i$ differ - if there are no such pairs of inputs, then by induction one can show that $f$ is IIA. W.l.o.g, $f_j(\mathbf{y}) = y_j^i \neq f_j(\mathbf{x})$, so $y^i$ is a partial manipulation of $x$.

|  | Outputs agree on issue $j$ | Outputs disagree on issue $j$ | |
|---|---|---|---|
| Inputs agree on issue $j$ | IIA | Anti-IIA | |
| | No partial manipulation | 1 directional partial manipulation | |
| Inputs disagree on issue $j$ | | Monotone | Anti- Monotone |
| | No partial manipulation | 2 directional partial anti manipulation | 2 directional partial manipulation |

Table 6: IIA and partial manipulations

∎

---

[5]The same theorem and proof hold for the genaral case where the codomain of $f$ is a bigger subset of $\{0,1\}^m$, i.e. $f : X^n \to Y$ and $X \subseteq Y \subseteq \{0,1\}^m$

The notion of IIA aggregators is well studied ([NP10, DH10]), and there is a full characterization of the evaluation spaces $X$ for which there is an impossibility theorem. The main property in this context is called *Totally Blocked*, which we will not define here.

The impossibility theorem for PMF aggregators is:

**Theorem 3.2:** *([NP10]) Every monotone and IIA aggregator $f : X^n \to X$ is dictatorial, if and only if an evaluation space $X \subseteq \{0,1\}^m$ is Totally Blocked.*

This theorem, combined with theorem 3.1 yields the following characterization of the cases for which there exists a a non-dictaroial PMF-aggregator:

**Corollary 3.3:** *([NP10]) Every PMF aggregator $f : X^n \to X$ is dictatorial, if and only if an evaluation space $X \subseteq \{0,1\}^m$ is Totally Blocked.*

### 3.2.1 The connection to social choice

As mentioned in the introduction, manipulation in the social choice setting is a situation when a voter reports a false ranking and the resulting alternative chosen is better w.r.t. her original ranking - $f(x^i, x^{-i}) <_{x^i} f(y, x^{-i})$

**Theorem 3.4:** *([Gib73, Sat75]) For $m \geq 3$, every social choice function $f : L_m^n \to [m]$ that is onto and strategy-proof, is dictatorial.*

There are two differences between the setting of social choice and judgment aggregation. Those are the definition of manipulation, and the fact that in social choice, the output of the social aggregator is not of the same space of the inputs.

The setting of judgment aggregation can be generalized to include the case of social choice as a private case, by allowing the ouputs of the aggregation mechanism to be partial evaluations, in which not all issues are necessarily decided. We shall call this setting *Partial Judgment Aggregation* (PJA). We define here the appropriate analogue of Partial manipulations for PJA. The manipulations in the social choice setting (used in the Gibbard-Satterthwaite theorem) identify with this definition. However, it is not straightforward to generalize the proofs shown in the previous subsection and obtain GS from Arrow's theorem, and we shall not show it here.

The definitions of preference and partial evaluations need to be modified only slightly. Again, let $f$ be an aggregator, $\mathbf{x} \in X^n$ a profile, $z = f(x^i, x^{-i})$ the aggregated evaluation, $y$ a false evaluation and $w = f(y, x^{-i})$ the resulting manipulated aggregated evaluation. For an issue $j \in J$, **if $z_j$ is decided** and $w_j = x^i_j \neq z_j$, we shall call $w$ *j-preferrable* over $z$ according to $x^i$.

**Definition 3.5: Partial Manipulation in PJA:** If there exists an issue $j \in J$ such that $w$ is $j$ preferrable over $z$ according to $x^i$, then $y$ is a *partial manipulation* of $i$ over $\mathbf{x}$. ∎

It is possible to define a limited analogue of IIA for PJA and follow the proof of theorem 3.1 to prove its analogue for PJA, but we will not show it here.

# 4 Full manipulation

## 4.1 Motivation

As was shown, designing an aggregator that is immune to partial manipulations is not always possible. In that case, we may still like to prevent weaker types of manipulation, with the weakest being full manipulation. More over, a manipulation-free aggregator under any type of definition is also full-manipulation free. Therefore, understanding the space of full-manipulation-free (FMF) aggregators is helpful in the design of manipulation-free aggregators under other definitions.

In this section we describe a set of aggregators which are FMF and also minimalize the cases in which there is a partial manipulation.

The natural question that comes up is what are the conditions on $X$ such that there exists a FMF aggregator that is not dictatorial. It turns outs that for any set set $X$ there are such functions. We shall show an easy construction of aggregators that are FMF and not strictly dictatorial, but are still very far from being anonymous.

However, such aggregators are not as interesting, as the number of influential voters in such a scheme is independent on $n$. We shall focus more on the construction of an annonymous FMF aggregator, and show that it is also possible for every evaluation space $X$.

## 4.2 Partitions of issues

We design a family of non dictatorial FMF aggregators based on a partition of the set of issues $J$ to the set of voters, called *partition aggregators*. Consider the folowing partition of $m$ issues into $n$ subsets, $K = K_1 \cup K_2 \cup ... \cup K_n$ where $K_i \cap K_j = \emptyset$, (it is possible that some of the voters won't have any influence, i.e $K_i = \emptyset$) W.l.o.g we will assume that $K_1 = \{1, 2...t_1\}, K_2 = \{t_1+1, ..., t_2\}, ...K_n = \{t_{n-1}+1...t_n\}$. We go over the issues sequentially. The decision on issue $i \in K_j$ will follow the opinion of voter $j$ unless the resulting partial evaluation on the issues $1, ..., i-1, i$ is infeasible. Formally, we define the social aggregator $f : X^n \to X$ inductively over $i$ going from 1 to $m$ to be:
$$(f(x)_i)_{i \in K_1} = x_i^1$$
and
$$(f(x)_i)_{i \in K_j} = \begin{cases} x_i^j & \text{the partial evaluation } (f(x)_1, ...f(x)_{i-1}, x_i^j) \text{ is feasible} \\ 1 - x_i^j & \text{otherwise} \end{cases}$$

The aggregator is consistent as a result of the inductive construction. The aggregator is a FMF aggregator since an agent $j$ can change the result on issue $i$ only by changing the result in at least one other issue in which his opinion was accepted. Therefore we get the following proposition

**Proposition 4.1:** *For every $X \subseteq \{0,1\}^k$, any partition aggregator is a FMF aggregator.*

A particularly interesting example an *almost dictatorial* aggregator, obtained by taking $K_1 = \{1, ..., m-1\}, K_2 = \{m\}$

$$f(x) = \begin{cases} (x_1^1, ..., x_{m-1}^1, x_m^2) & (x_1^1, ..., x_{m-1}^1, x_m^2) \in X \\ x^1 & otherwise \end{cases}$$

Note that the almost dictatorial aggregator is non manipulable, not only for this weak definition, but also for the weighted Hamming definition, for every $X$, when the issue determined by the second voter is the issue with the minimal weight. This means that there can be no impossibility theorem in the flavour of GS for the weighted Hamming manipulation.

Of course, it is not necessarily PMF for every $X$. there can be cases where it is beneficial for the voter deciding on the first $m-1$ issues to lie in order to gain on the $m$'th issue, by denying the second voter his influence.

## 4.3 Annonymous FMF Aggregators

An important approach for designing FMF aggregators is based on PMF aggregators. A basic property for a society would have to be to avoid partial manipulations whenever it is possible. From 3.1 we get that an aggregator is PMF if and only if is IIA and monotone. This fact is true not only for a consistent aggregator $f : X^n \to X$ but also for an aggregator from $X^n \to \{0,1\}^m$. As we saw in the impossibility theorem an IIA and monotone aggregator does not always produce outputs consistent with the evaluation space $X$. Therfore, we would like to correct these functions in the places where they are not consistent. We would like to study the set of aggregators which are consistent and yet "close" to an IIA and Monotononic aggregator.

Formally, for an incosistent function $g : X^n \to \{0,1\}^m$, a consistent function $f : X^n \to X$ is called a correction of $g$ if $f(x) = g(x)$ whenever $g(x)$ is consistent. Denote by $\mathbb{M}$ the set of all IIA and monotone functions $f : X^n \to \{0,1\}^m$. when $f$ is a correction of a function $m \in \mathbb{M}$, at least all pairs of inputs for which $m$ falls into $X$ do not form a partial manipulation. We shall denote by $\mathbb{F}$ the set of consistent functions which are a correction of a function in $\mathbb{M}$. The functions in $\mathbb{F}$ will be called *close to partial manipulation free* aggregators (C-PMF).

Our aim in this chapter is to build a FMF-aggregator $f$ with the property of being a C-PMF aggregator. We shall define the subset $\mathbb{G}$ of $\mathbb{F}$ to be the set of functions who are a composition of a function $g : \{0,1\}^m \to X$ with a function $m \in \mathbb{M}$ such that for every feasible evaluation $x \in X$, $g(x) = x$. Being a member of $\mathbb{G}$ means that the 'correction' part of the social aggregator in the cases where $m$, the IIA and Monotone stage, is not consistent, depends only on the outcome of $m$ and not on the entire on the whole profile.

A special subset $\mathbb{H}$ of $\mathbb{G}$ is a composition of a Hamming nearest neighbour function $h$ with a function $m \in \mathbb{M}$. A Hamming nearest neighbour function $h : \{0,1\}^m \to X$ is a function that, given $x \in \{0,1\}^m$, returns a closest point in $X$, under a given Hamming metric, i.e each issue has a nonzero weight[6]. Of course, such a function is not properly defined without a tie-breaking rule. We need to set proper tie-breaking rules in order to avoid manipulations. The main property we wish to maintain is that, given a nearest neighbour function $h$, if two different points $a, b \notin X$ both have the points $\alpha, \beta \in X$ in their set of potential nearest neighbours according to $h$, then it can not be that $h(a) = \alpha$ and $h(b) = \beta$.

---

[6]By definition any metric must maintain the following properties: non-negativity, idendity of indiscernibles, symmetry and the triangle inequalility. It is easy to check that a weighted Hamming distance is a metric if and only if each issue has a nonzero weight.

One way of implementing that property is by choosing according to some lexicographical order in case of a tie. We shall denote the set of functions using the lexicographical tie-breaker as $\mathbb{H}^1$ and the set of functions satisfying the aforementioned property as $\mathbb{H}^2$, so:

$$\mathbb{H}^1 \subseteq \mathbb{H}^2 \subseteq \mathbb{H} \subseteq \mathbb{G} \subseteq \mathbb{F}$$

. We shall use the following notation in order to present the geometric relations of binary vectors $a, b, c$. We say that $c$ is *between* $a, b$ if for every coordinate $i$ $a_i \leq c_i \leq b_i$ or $b_i \leq c_i \leq a_i$. The notation $[a, b]$ will describe the set of all the vectors between a and b $[a, b] = \{v | if\ a_i = b_i\ than\ v_i = a_i\}$. Likewise, $(a, b)$ describes the set $[a, b] \backslash \{a, b\}$ and $[a, b) = [a, b] \backslash \{b\}$, etc. We say that $a \in X$ is a *neighbour* of $b \notin X$ if $(a, b) \cap X = \emptyset$

**Theorem 4.2:** *For every $X \subseteq \{0,1\}^k$, any social aggregator $f = h \circ m \in \mathbb{H}^2$ ($m \in \mathbb{M}$), $f$ is a FMF aggregator. Furthermore, if $m$ is annonymous, then $f$ is annonymous.*

**Proof:**

Assume we have a profile $\mathbf{x} \in X^n$ and that voter $i$ tries to manipulate by reporting his opinion to be $y^i \in X$ instead of $x^i$ (denote the profile $(y^i, x^{-i})$ as $\mathbf{y}$). Since $m$ is IIA and monotone, it is easy ot see that $m(\mathbf{x})$ is between $x^i$ and $m(\mathbf{y})$. Therefore, the following is a full partition of the set of issues $J$:

$$A_1 = \{j | x_j^i = m(x^i, x^{-i})_j = m(y, x^{-i})_j\}$$
$$A_2 = \{j | x_j^i = m(x^i, x^{-i})_j \neq m(y, x^{-i})_j\}$$
$$A_3 = \{j | x_j^i \neq m(x^i, x^{-i})_j = m(y, x^{-i})_j\}$$

We shall further partition each of these subsets $A_t$ according to their relation to $f(\mathbf{x})$ and $f(\mathbf{y})$.

$$A_{t1} = \{j \in A_t | x_j^i = f(\mathbf{x})_j = f(\mathbf{y})_j\}$$
$$A_{t2} = \{j \in A_t | x_j^i = f(\mathbf{x})_j \neq f(\mathbf{y})_j\}$$
$$A_{t3} = \{j \in A_t | x_j^i = f(\mathbf{y})_j \neq f(\mathbf{x})_j\}$$
$$A_{t4} = \{j \in A_t | x_j^i \neq f(\mathbf{x})_j = f(\mathbf{y})_j\}$$

Recall that $f(\mathbf{x})$ and $f(\mathbf{y})$ are nearest neighbours of $m(\mathbf{x})$ and $m(\mathbf{y})$, respectively and w.l.o.g. assume that $x^i = \vec{1}$. Table 7 summarizes the relevant evaluations under the partition we defined.

In order to prove that judge $i$ does not gain a full manipulation by choosing $y$ instead of $x^i$ we have to show that the set of columns on which $f(\mathbf{y})$ and $x^i$ differ is not a proper subset of the set of columns on which $f(\mathbf{x})$ and $x^i$ differ. Therefore we have to show that for any set $X$ and any profile $\mathbf{x}$ and any judge $i$

$$(1) \qquad A_{12} \cup A_{22} \cup A_{32} \neq \emptyset$$

|       | $A_{11}$ | $A_{12}$ | $A_{13}$ | $A_{14}$ | $A_{21}$ | $A_{22}$ | $A_{23}$ | $A_{24}$ | $A_{31}$ | $A_{32}$ | $A_{33}$ | $A_{34}$ |
|-------|------|------|------|------|------|------|------|------|------|------|------|------|
| $x_i$ | 1 | 1 | 1 | 1 | 1 | 1 | 1 | 1 | 1 | 1 | 1 | 1 |
| $m(\mathbf{x})$ | 1 | 1 | 1 | 1 | 1 | 1 | 1 | 1 | 0 | 0 | 0 | 0 |
| $m(\mathbf{y})$ | 1 | 1 | 1 | 1 | 0 | 0 | 0 | 0 | 0 | 0 | 0 | 0 |
| $f(\mathbf{x})$ | 1 | 1 | 0 | 0 | 1 | 1 | 0 | 0 | 1 | 1 | 0 | 0 |
| $f(\mathbf{y})$ | 1 | 0 | 1 | 0 | 1 | 0 | 1 | 0 | 1 | 0 | 1 | 0 |

Table 7: $\mathbb{H}^2$ - the general case

Assume by contradiction that $A_{12} = A_{22} = A_{32} = \emptyset$. Since $f(\mathbf{x})$ is closer (by the hamming distance) than $f(\mathbf{y})$ to $m(\mathbf{x})$ we get that

$$(2) \quad |A_{13}| + |A_{23}| + |A_{32}| \leq |A_{12}| + |A_{22}| + |A_{33}|$$

where $|A| \equiv \Sigma_{i \in A} \omega_i$. On the other hand, by the fact that $f(\mathbf{y})$ is closer than $f(\mathbf{x})$ to $m(\mathbf{y})$ we get that

$$(3) \quad |A_{12}| + |A_{23}| + |A_{33}| \leq |A_{13}| + |A_{22}| + |A_{32}|$$

By combining the two inequlities together we get that:

$$(4) \quad |A_{23}| \leq |A_{22}|$$

Therefore $A_{12} = A_{22} = A_{23} = A_{32} = \emptyset$, and we get from inequality (2) and (3) that

$$|A_{13}| \leq_{(2)} |A_{33}| \leq_{(3)} |A_{13}|.$$

We conclude that both $f(\mathbf{x})$ and $f(\mathbf{y})$ are in the set of nearest neighbours of $m(\mathbf{x})$ and $m(\mathbf{y})$ and since $h \in \mathbb{H}^2$ we must have $f(\mathbf{x}) = f(\mathbf{y})$, so there is no manipulation (particularly, if $h \in \mathbb{H}^1$ we can't get $|A_{13}| = |A_{33}|$ unless they are equal to 0). ∎

Theorem 4.2 does not hold for any function in $\mathbb{F}$. Even if we use a function in $\mathbb{G}$ and the correction is done by choosing a neighbour which is not necessarily a nearest neighbour, then the aggregator is not necessarily FMF.

### 4.3.1 Social welfare maximizer

An important concept in mechanism design in the *social welfare maximizer*. Each individual in the society has a function returning his welfare given his opinion and the aggregated opinion. A social welfare maximizer is an aggregator that always returns the evaluation that maximizes the total welfare of all individuals in the society.

We consider the case where the welfare of every individual $i$ is $-d_\omega(x^i, f(\mathbf{x}))$, the Hamming distance between his opinion and society's opinion, according to some weight function with positive weights on all the issues. The corresponding social welfare maximizer is the function

$$f(\mathbf{x}) = argmin_{x \in X} \sum_{i \in N} d_\omega(x, x^i)$$

We call this aggregator $f$ a Hamming social welfare maximizer. In case of a tie, we shall use a tie-breaking rule which will ensure that $f \in \mathbb{H}^2$. Notice that $f \in \mathbb{F}$ since it is the correction of the IIA and monotone aggregator $\tilde{f} : X^n \to \{0,1\}^k$ where

$$\tilde{f}(\mathbf{x}) = argmin_{x \in \{0,1\}^k} \sum_{i \in N} d_\omega(x, x^i)$$

However, $f \notin \mathbb{G}$, because the correction depends on the entire propile. It is easy to construct two profiles $\mathbf{x}, \mathbf{y}$ such that $\tilde{f}(\mathbf{x}) = \tilde{f}(\mathbf{y})$ and $f(\mathbf{x}) \neq f(\mathbf{y})$ [7].

We shall show that this aggregtor has the same property of being FMF.

**Theorem 4.3:** *For every evaluation space $X \subseteq \{0,1\}^k$, a Hamming social welfare maximizer is FMF and annonymous.*

**Proof:** Let $\mathbf{x} = (x^i, x^{-i})$ be the profile with i's opinion, and $\mathbf{y} = (y, x^{-i})$ with i's declaration[8]. W.l.o.g. $x^i = \vec{1}$. We partition the issues into sets in a simialr fashion to theorem 4.2, as shown in table 8.

|       | $A_{11}$ | $A_{12}$ | $A_{21}$ | $A_{22}$ | $A_{31}$ | $A_{32}$ | $A_{41}$ | $A_{42}$ |
|-------|----------|----------|----------|----------|----------|----------|----------|----------|
| $x^i$ | 1        | 1        | 1        | 1        | 1        | 1        | 1        | 1        |
| $y$   | 1        | 0        | 1        | 0        | 1        | 0        | 1        | 0        |
| $f(\mathbf{x})$ | 1 | 1 | 1 | 1 | 0 | 0 | 0 | 0 |
| $f(\mathbf{y})$ | 1 | 1 | 0 | 0 | 1 | 1 | 0 | 0 |

Table 8: Nearest neighbour - social welfare maximaizer

Acording to profile $\mathbf{x}$, profile $f(\mathbf{x})$ is prefered over profile $f(\mathbf{y})$ i.e.

$$\sum_{k \in N \setminus \{i\}} d_\omega(f(\mathbf{x}), x^k) + d_\omega(f(\mathbf{x}), x^i) \leq \sum_{k \in N \setminus \{i\}} d_\omega(f(\mathbf{y}), x^k) + d_\omega(f(\mathbf{y}), x^i)$$

And:

$$\sum_{k \in N \setminus \{i\}} d_\omega(f(\mathbf{y}), x^k) + d_\omega(f(\mathbf{y}), y) \leq \sum_{k \in N \setminus \{i\}} d_\omega(f(\mathbf{x}), x^k) + d_\omega(f(\mathbf{x}), y)$$

And by combining the two inequalities we get that

$$d_\omega(f(\mathbf{y}), y) - d_\omega(f(\mathbf{x}), y) \leq \sum_{k \in N \setminus \{i\}} d_\omega(f(\mathbf{x}), x^k) - \sum_{k \in N \setminus \{i\}} d_\omega(f(\mathbf{y}), x^k) \leq d_\omega(f(\mathbf{y}), x^i) - d_\omega(f(\mathbf{x}), x^i)$$

Because of the tie-breaking rules we use, at least one of the inequalities must be strong. By focusing on the entries $A_{21}, A_{22}, A_{31}, A_{32}$ where $f(\mathbf{x})$ and $f(\mathbf{y})$ differ we get:

$$\sum_{j \in A_{21} \cup A_{32}} \omega_j - \sum_{j \in A_{22} \cup A_{31}} \omega_j < \sum_{j \in A_{21} \cup A_{22}} \omega_j - \sum_{j \in A_{31} \cup A_{32}} \omega_j$$

---

[7] For example let $X = \{110000, 001000, 000111\} \subseteq \{0,1\}^6$ and $n = 9$. $\mathbf{x}$ will be the profile where 3 agents hold the first opinion 110000, 2 agents hold the second opinion 001000 and 4 agents hold the last opinion 000111. $\mathbf{y}$ will be the profile where 3 agents hold the first opinion,3 agents hold the second one and 3 agents hold the last one. By taking the uniform weights we get that $f(\mathbf{x}) = 000111$, $f(\mathbf{y}) = 001000$ and $\tilde{f}(\mathbf{x}) = \tilde{f}(\mathbf{y}) = 000000$

[8] Denote that in contrast to Theorem 4.2 it maybe that $\tilde{f}(\mathbf{x}) = \tilde{f}(\mathbf{y})$.

And therfore:
$$\sum_{j \in A_{32}} \omega_j < \sum_{j \in A_{22}} \omega_j$$

If the aggregator is not FMF it must be that $A_{21} \cup A_{22} = \emptyset$ and $A_{31} \cup A_{32} \neq \emptyset$ but it is contrary to the last inequality. ∎

The Hamming social welfare maximizer has been used before. In preference aggregation, it is known as Kemeny's rule ([LY78]). There are many works that discuss various aspects of Kemeny's rule[9], but not in connection with strategy-proofness, as far as we know. facility location [NAT10], classfication [RMR09] and more [Pig06]. A general connection between social welfare maximization and strategy-proofness was not previously known.

## 5 Hamming manipulations

### 5.1 Main Results

In this section we present some results regarding the Hamming manipulation. We say that voter $i$ with opinion $x$ prefers the result $v \in X$ more than $u \in X$ if the distance, according to a weighted Hamming metric $\omega$, $d_\omega(x,v)$ of $v$ from $x$ is less than the distance of $u$ from $x$.

As was mentioned in the previous chapter, the almost dictator aggregator is HMF for any weighted hamming definition. Therefore, we will focus on the interesting case of building an annonymous HMF aggregator. Following the results of the previous chapter, we focus on the set of aggregators $\mathbb{H}$. We show two conditions for determining whether an aggregator $f \in \mathbb{H}$ is not only FMF, but also HMF. We shall discuss the cases in which such an aggregator is non-HMF. More over, we use these two lemmas to analyze some special cases and show whether there is a HMF aggregator in $\mathbb{H}$.

We show that in any case where there is a manipulation of an aggregator $h \circ m \in \mathbb{H}$ on the profile $\mathbf{x}$, the 2 intermediate results $w = m(x^i, x^{-i})$ and $z = m(y, x^{-i})$ must both be outside of $X$, not too "far" from each other (lemma 5.4) and not too "close" to each other (lemma 5.5). For that we will use a combinatorial representation of the evaluation space $X$.

**Definition 5.1:** For a non-empty evaluation space $X \subseteq \{0,1\}^m$, A minimally infeasible partial evaluation (abbreviated MIPE) is a K-evaluation $x = (x_j)_{j \in K}$ for some $K \subseteq J$ which is infeasible, but such that every restriction of $\mathbf{x}$ to a proper subset of $K$ is feasible. ∎

$X$ can be defined by its set of MIPEs. A MIPE represents a maximal Boolean subcube that is outside of $X$[10].

**Definition 5.2:** For every MIPE $a = (a_j)_{j \in K}$ we denote by $T_a$, the *MIPE-set* of $a$, as the following subset of $\{0,1\}^m$: $T_a = \{x | x_{|K} = a\}$. ∎

---
[9]see, for example [NAN08]

[10]An IIA and monotone aggregator over $X$ is annonymous, neutral, PMF and consistent iff all its MIPES are of size 2 [NP10]

**Definition 5.3:** For every evaluation $x \in X^c$, we denote by $MT(x)$, its *MIPE-type* as the following set of MIPES of $X$: $MT(x) = \{a | x \in T_a\}$ ∎

We are now prepared to bring the partial characterizations for the general case:

**Lemma 5.4:** *For every $X \subseteq \{0,1\}^k$, and any social aggregator $f \in \mathbb{H}$, $f = h_\omega \circ m$, if $y$ is a $\omega$-Hamming manipulation of $i$ over $(x^i, x^{-i})$, then $[(m(x^i, x^{-i})), (m(y, x^{-i}))] \cap X = \emptyset$. (In other words there exists a MIPE $a$ such that $(m(x^i, x^{-i})), (m(y, x^{-i})) \in T_a$.)*

**Proof:** Let $m(x^i, x^{-i}) = v$, and $m(y, x^{-i}) = u$. We shall show that if $[(m(x^i, x^{-i})), (m(y, x^{-i}))] \cap X \neq \emptyset$ then $d_\omega(x^i, h_\omega(v)) \leq d_\omega(x^i, h_\omega(u))$, and therefore $y$ is not a succesful Hamming manipulation. (For convenience sake, we shall drop the $\omega$ subscript from now on)

By the Triangle inequality we get

$$(1) \quad d(x^i, h(u)) \geq d(x^i, u) - d(h(u), u)$$

And

$$(2) \quad d(x^i, h(v)) \leq d(x^i, v) + d(h(v), v)$$

Let $t$ be any point in $[v, u] \cap X$. Since $h$ is a Hamming nearest neighbour function, we get that:

$$(3) \quad d(x^i, h(u)) \geq d(x^i, u) - d(t, u)$$

And

$$(4) \quad d(x^i, h(v)) \leq d(x^i, v) + d(t, v)$$

Since $m$ is monotone, $v$ is on a shortest path from $x^i$ to $u$. Also, since $t \in [v, u]$, $t$ is on a shortest path between $v$ and $u$. Therefore,

$$(5) \quad d(x^i, v) + d(t, v) = d(x^i, t) = d(x^i, u) - d(t, u)$$

Combining $(3), (4), (5)$ we get:
$$d(x^i, h(u)) \geq d(x^i, h(v))$$

∎

**Lemma 5.5:** *For every $X \subseteq \{0,1\}^k$, and any social aggregator $f \in \mathbb{H}$ $f = h_\omega \circ m$, if $y$ is a $\omega$-Hamming manipulation of $i$ over $(x^i, x^{-i})$, then $MT(m(x^i, x^{-i})) \neq MT(m(y, x^{-i}))$.*

**Proof:** Let $m(x^i, x^{-i}) = v$, and $m(y, x^{-i}) = u$ and assume $MT(v) = MT(u)$. We shall show that agent $i$ can't manipulate by declaring $y$ instead of $x$. Assume that $v = u + 1_S$ where $1_S$ is the indicator of the subset $S \subseteq J$. For every issue $j \in J$, $v_j = u_j$ if and only if $j \notin S$. $v$ and $u$ are in the same MIPE-sets and therefore we get that for every MIPE-set $T_{a_K}$ such that $v, u \in T_{a_K}$ we have $S \cap K = \emptyset$. And therefore:

$$S \cap \left( \bigcup_{a_{K_i} \in MT(v)} K_i \right) = \emptyset$$

Claim: $v + 1_W$ is a neighbor of $v$ if and only if $u + 1_W$ is a neighbor of $u$.
Proof: $v + 1_W$ is a neighbor of $v$ if and only if for every set $K_i$ such that $a_{K_i} \in MT(v)$ we have that $K_i \cap W \neq \emptyset$ and for every nonempty subset $W'$ of $W$ there exist a set $a_{K_i} \in MT(v)$ such that $K_i \cap W' = \emptyset$. In other words, $W$ is a set cover of $K_1, K_2...K_t$ that is minimal to inclusion. Since $MT(v) = MT(u)$ the result follows.

More over, as a result of the fact that $W$ is a *minimal* set cover, we get that

$$W \subseteq \bigcup_{a_{K_i} \in MT(v)} K_i$$

and therefore:

$$W \cap S = \emptyset$$

Assume the nearest neighbour of $v$ is $h_\omega(v) = v + 1_W$, then the claim implies that the nearest neighbour of $u$ is $h_\omega(u) = u + 1_W = v + 1_S + 1_W = h_\omega(v) + 1_S$. Again, since $m$ is monotone, $v$ agrees with $x_i$ on $S$ and $u$ disagrees with $x_i$ on $S$. Since $W \cap S = \emptyset$, then also $h(v)$ agrees with $x_i$ on $S$ and $h(u)$ disagrees with $x_i$ on $S$. Therefore

$$d(x^i, h_\omega(u)) = d(x^i, h_\omega(v)) + |S|$$

and $h(u)$ is further away from $x_i$ than $h(v)$. ∎

## 5.2 examples

These two theorems do not give a full characterization for the sets $X$ for which there exists a manipulation free aggregator. However, they show that for aggregators in $\mathbb{H}$, a Hamming manipulation occurs only in special circumstances. For many particular cases, including the preference aggregation model, we can conclude whether or not there exists an HMF aggregator in $\mathbb{H}$. In this subsection We shall present for two particular cases[11] the preference model and the "k choose m" model, to be defined later on.

We shall show that for the preference aggregation model, when there are three alternatives any combination of a monotone aggregator and the standard nearest neighbor aggregator is a HMF aggregator but not for more than three alternatives. A general natural question that arises (and is still open) is what is the minimal number of alternatives for which there is no anonymous HMF and C-PMF aggregator.

For the "k choose m" decision example we shall present some anonymous HMF C-PMF aggregators for any number $k$ and $m$. Those examples will give us some intuition regarding the existence of HMF aggregators and the usage of the Theorems.

### 5.2.1 preference aggregation

We shall denote the set of alternatives by $A = \{a, b, c, ...\}$, $|A| = k$. The set of issues $K$ is the set of pairwise preferences between every 2 alternatives, so $m = \binom{k}{2}$. For $k > 2$ it is well known from Arrow's theorem that there is no IIA and Monotone aggregator and therefore there isn't a PMF aggregator. In the next claim we shall show that there is an anonymous, HMF and C-PMF aggregator for three alternatives.

---

[11] The cases of facility location on a line and a cycle are shown in a subsequent work.

**Claim 5.6:** If $m = 3$, then all aggregators $h_\omega \circ m$ in $\mathbb{H}$ are $\omega$-HMF.

**Proof:** For $k = 3$, the set of issues is $\{a \succ b, b \succ c, c \succ a\}$, and the permissible evaluations are all possible evaluations except for 000 and 111, which encode a non-transitive order. According to theorem 5.4 we immediately get that a manipulation can't occur. A manipulation can 0ccur only if $m(x^i, x^{-i})$ and $m(y, x^{-i})$ are different and $[(m(x^i, x^{-i})), (m(y, x^{-i}))] \cap X = \emptyset$. ∎

For more than three alternatives we will not bring a full answer to the question of whether there exists an anonymous HMF and C-PMF aggregators and we will show that it can't be of the form $h_\omega \circ m$. [12]

**Claim 5.7:** In preference aggregation over at least $k \geq 4$ alternatives, and at least 3 voters, aggregators $h \circ maj \in \mathbb{H}$ are not HMF.

**Proof:** According to theorem 5.5, in order to build two profiles in which there is a manipulation we would like to have two evaluations $v_1, v_2 \notin X$ and two MIPES $w_1, w_2$ (all the MIPES in this example are of leghth 3 and present a cycle containing three alternatives) such that $v_1, v_2 \in T_{w_1}$ and exactly one of them is in $T_{w_2}$. More over, an infeasible evaluation appears in more than one MIPE iff it presents a cycle of four alternatives. Any such evaluation has **only** one neighbor of distance one, therefore it has only one nearest neighbor.

We demonstrate the possibility of constructing a manipulation via the following example:

|  |  | $a \succ b$ | $b \succ c$ | $c \succ a$ | $a \succ d$ | $b \succ d$ | $c \succ d$ |
|---|---|---|---|---|---|---|---|
| Judge 1 | $a \succ b \succ d \succ c$ | 1 | 1 | 0 | 1 | 1 | 0 |
| Judge 2 | $b \succ c \succ a \succ d$ | 0 | 1 | 1 | 1 | 1 | 1 |
| Judge 3 | $d \succ c \succ a \succ b$ | 1 | 0 | 1 | 0 | 0 | 0 |
| Maj | $(abc), (acd), (abdc)$ | 1 | 1 | 1 | 1 | 1 | 0 |
| HNN aggregator | $a \succ b \succ d \succ c$ | 1 | 1 | **0** | 1 | 1 | 0 |

Table 9: Preference aggregation - four candidates

If judge 2 gives a different opinion on issue $a \succ d$ we get a manipulation, as shown in table 10. *

It is easy to see that this example is general in the sense that for any odd number of judges and any aggregator that chooses the nearest neighbour it is possible to construct such an example with the same result in the majority vote stage and where judge 2 is pivotal in the column $a \succ d$ ∎

---
[12]In another work in which we use random functions it can be shown that there exists an HMF aggregator $h_\omega \circ m$ for four alternatives, where $h$ is random and $m$ is monotone

|               |                           | $a \succ b$ | $b \succ c$ | $c \succ a$ | $a \succ d$ | $b \succ d$ | $c \succ d$ |
|---------------|---------------------------|-------------|-------------|-------------|-------------|-------------|-------------|
| Judge 1       | $a \succ b \succ d \succ c$ | 1 | 1 | 0 | 1 | 1 | 0 |
| Judge 2       | $b \succ c \succ d \succ a$ | 0 | 1 | 1 | **0** | 1 | 1 |
| Judge 3       | $d \succ c \succ a \succ b$ | 1 | 0 | 1 | 0 | 0 | 0 |
| Maj           | $(abc), (abd), (abdc)$    | 1 | 1 | 1 | **0** | 1 | 0 |
| HNN aggregator | $b \succ d \succ c \succ a$ | **0** | 1 | 1 | 0 | 1 | 0 |

Table 10: Manipulation of the second judge

### 5.2.2 $\binom{k}{m}$-decision

One special example is the following, in which a society would like to choose $k$ candidates out of $m$ possible candidates. Formally, $X \subseteq \{0,1\}^m, X = \{x \mid |x| = k\}$. We wish to design a a C-PMF and HMF aggregator $f : X^n \to X$, when the weights over the issues are uniform.

**Claim 5.8:** The social welfare maximizer under a lexical tie breaking rule is C-PMF and HMF

**Proof:**

Since the size of the evaluations and the outcome is fixed, we get that, the SWM returns a vector $z \in X$ such that $z = argmax_z\{\sum_i <x^i, z>\}$, because

$$SWM(x) = argmax_z \sum_i -d(x^i, z) = argmax_z \sum_i -|x^i - z|_2^2 =$$

$$argmax_z \sum_i - \left(|x^i|_2^2 + |z|_2^2 - 2 <x^i, z>\right) = argmax_z \sum_i <x^i, z> -2k$$

Therefore, the SWM is a C-PMF, because

$$\sum_i <x^i, z> = \sum_i \sum_j x_j^i \cdot z_j = \sum_j \sum_i x_j^i \cdot z_j = \sum_j z_j |x_j|$$

so the SWM returns the $k$ highest candidates ordered first by their number of votes and then by some predetermined tie breaking order. Take an IIA function that applies a certain threshold function to each of the columns $x_j$. If it is consistent, it means that exactly $k$ candidates passed the threshold. Therefore, the SWM will also pick these $k$ candidates, so the SWM agrees with this IIA and monotone function.

We shall now show that it is an HMF. Assume, by contradiction, that there exists $x, i$ and $y$ such that $y$ is a Hamming manipulation of $i$ over $x$, and let $z = SWM(x), w = SWM(y, x^{-i})$. A simple argument shows that we also have $SWM(w, x^{-i}) = w$. Indeed, for every possible outcome $v$ and evaluation $y$, we have

$$<w, w> - <v, w> = |w \cap v^c|$$

$$<w, y> - <v, y> \geq |w \cap v^c \cap y| \Rightarrow$$

$$<w,w> - <v,w> \geq <w,y> - <v,y>$$

Therefore, since
$$w = argmax_v \sum_{l \neq i} <x^l, v> + <y, v>$$

we also have
$$w = argmax_v \sum_{l \neq i} <x^l, v> + <w, v>$$

Define the sets:
$$U_0 = \{j | x_j^i = 0, z_j = 1, w_j = 0\}$$
$$U_1 = \{j | x_j^i = 1, z_j = 0, w_j = 1\}$$
$$D_0 = \{j | x_j^i = 1, z_j = 1, w_j = 0\}$$
$$D_1 = \{j | x_j^i = 0, z_j = 0, w_j = 1\}$$

and denote their sizes $u_0 = |U_0|, u_1 = |U_1|, d_0 = |D_0|, d_1 = |D_1|$.

Since $|z| = |w|$, we have
$$u_0 + d_0 = u_1 + d_1 \Rightarrow (u_0 - d_1) = (u_1 - d_0)$$

Since $w$ is a manipulation, we have
$$u_0 + u_1 > d_0 + d_1 \Rightarrow (u_0 - d_1) + (u_1 - d_0) > 0$$

Combining these we get that both $u_0 > 0$ and $u_1 > 0$. However, when $i$ voted $x^i$, the elements of $U_1$ were ranked below the elements of $U_0$. When $i$ voted $w$, he did not change his vote on the elements of $U_0$ and $U_1$. Therefore, the elements of $U_1$ could not have passed over the elements of $U_0$ in the order that determines the $k$ winners. ∎